\documentclass{ws-ijmpa}
\usepackage[super,compress]{cite}
\usepackage{graphicx}
\usepackage{hyperref}
\usepackage{slashed}

\newcommand{\beq}{\begin{eqnarray}}
\newcommand{\eeq}{\end{eqnarray}}

\newcommand{\nn}{\nonumber}

\DeclareRobustCommand{\eq}[1]{Eq.~(\ref{eq:#1})}

\DeclareRobustCommand{\sec}[1]{Sec.~\ref{sec:#1}}
\DeclareRobustCommand{\secs}[2]{Secs.~\ref{sec:#1} and \ref{sec:#2}}
\DeclareRobustCommand{\fig}[1]{Fig.~\ref{fig:#1}}

\begin{document}
	\markboth{Yong Zhao}{Unravelling High-Energy Hadron Structures with Lattice QCD}
	
	%
	\catchline{}{}{}{}{}
	%
	
	\title{Unravelling High-Energy Hadron Structures with Lattice QCD~\footnote{This writing is based on a talk for the T-2 Theory Seminar at Los Alamos National Laboratory on Feb. 1, 2018.}}
	
	\author{Yong Zhao}
	\address{Center for Theoretical Physics, Massachusetts Institute of Technology, 77 Massachusetts Avenue\\
		 Cambrige, MA 02139, USA\\
		 yzhaoqcd@mit.edu}

	\maketitle
	
	\begin{history}
		\received{\today}
		\revised{\today}
	\end{history}
	
	\begin{abstract}
		Parton distribution functions are key quantities for us to understand the hadronic structures in high-energy scattering, but they are difficult to calculate from lattice QCD. Recent years have seen fast development of the large-momentum effective theory which allows extraction of the $x$-dependence of parton distribution functions from a quasi-parton distribution function that can be directly calculated on lattice. The extraction is based on a factorization formula for the quasi-parton distribution function that has been derived rigorously in perturbation theory. A systematic procedure that includes renormalization, perturbative matching, and power corrections has been established to calculate parton distribution functions. Latest progress from lattice QCD has shown promising signs that it will become an effective tool for calculating parton physics.
		
	\end{abstract}
	
	\ccode{PACS numbers:}
	
	\tableofcontents
	
	\section{Introduction}	
	
	Understanding the inner structure of hadrons has been one of the most important questions underlying our understanding of nature. However, it remains a challenging task to have a precise picture of the hadrons from both theory and experiment.
	Take the proton as an example. It is well known that the proton is a bound state of three valence quarks and has a size of about 1 fm or 1/(200 MeV). The structure of the proton is much more abundant if one increases the resolution to much smaller than 1 fm. Due to the vacuum fluctuation and non-Abelian gauge interaction in quantum chromodynamics (QCD), quark-antiquark pairs and gluons are created and annihilated at every instant.
	The proton is actually a system of infinite number of quarks and gluons that are moving relativistically with spin and orbital angular momentum. To measure the distribution and motion of the quarks and gluons inside the proton, one needs to increase the energy of the probe to much higher than 200 MeV, which could end up breaking the proton into pieces.

	The deep inelastic scattering (DIS) and proton-proton ($pp$) collision are the most important experiments of this kind to measure the quark and gluon structures of the proton. In DIS or $pp$ collision experiments, the probe is accelerated to a few GeV to TeV to collide with a proton which is at rest or also moving at large momentum. To the probe, the proton is moving at almost the speed of light, $v\sim c$. Due to Lorentz dilation effect, the quarks and gluons contract in the longitudinal direction. When the probe interacts with the quark or gluon inside the proton, it usually exchanges a hard momentum $Q$ within a short time $t\sim Q^{-1}$, which is much smaller than the time scale $\Lambda^{-1}_{\rm QCD}$ of the interactions between the quarks and gluons bound inside the proton. Therefore, we can approximate this hard process as the probe interacting with a free quark or gluon without the influence from the spectator quarks and gluons. As a result, the proton can be treated as a beam of free quarks and gluons with each carrying a certain fraction of its longitudinal momentum.
	
	This is Feynman's simple parton model~\cite{Feynman:1969ej} that was introduced to describe hadrons in DIS experiments in the late 1960s. The probability density of a parton carrying a given fraction of the longitudinal proton momentum is called the parton distribution function (PDF).	The PDFs not only provide structural information of the hadron in the longitudinal momentum space, but are also key inputs for making predictions in the Standard Model for colliders such as Tevatron and LHC. According to QCD factorization theorems~\cite{Collins:1989gx}, the cross section for a hard scattering process can be factorized as convolution of a partonic cross section and the PDF. The former can be analytically calculated in perturbative QCD, while the latter is a universal property of the hadron that is instrinsically infrared (IR) and nonperturbative. In the past half century, tremendous efforts have been made to extract the PDFs from high-energy experiments at state-of-the-art hadron physics programs including SLAC, DESY, FermiLab, CERN, Jefferson Lab, RHIC, etc. Currently, our best knowledge of the PDFs comes from the global analysis of cross section data ~\cite{Martin:2009iq,Buckley:2014ana,Ball:2014uwa,Dulat:2015mca,Alekhin:2017kpj}, which has achieved high precision for certain range of parton flavors and kinematics and made significant contributions to nuclear and particle physics such as the discovery of the Higgs boson~\cite{Aad:2012tfa,Chatrchyan:2012xdj}.
	
	The remarkable experimental achievements on the determination of PDFs have motivated theoretical efforts to calculate them from first principles. Compared to the exisiting experiments that are limited to certain parton flavors, spin strucutres and range of Bjorken-$x$~\cite{Agashe:2014kda}, one should expect the theoretical calculation to cover complementary---if not complete---information of the PDFs. If a systematic approach to calculate the PDFs is available, then the time and cost of improving the precision is likely to be shorter and less than buidling more advanced and powerful colliders.
	
	In modern theory, the PDFs are defined from the light-cone correlators of quarks and gluons in a hadron state. For example, the unpolarized quark PDF is
	\begin{align}\label{eq:lcpdf}
	q_i(x,\mu) \equiv \!\int\!\! \left.\frac{d\xi^-}{4\pi} \, e^{-ixP^+\xi^-} 
	\! \big\langle P \big| \bar{\psi_i} (\xi^-) \gamma^+ 
	W(\xi^-\!,0)
	\psi_i(0)\right|_{\mu} \big|P\big\rangle ,
	\end{align}
	where $x\in[0,1]$ is the momentum fraction of the parton, $i$ is a flavor index, the hadron momentum $P^\mu=(P^0,0,0,P^z)$, and $\xi^\pm = (t\pm z)/\sqrt{2}$ are the light-cone coordinates. The Wilson line $W$ is given by the path-ordered exponential
	\begin{align}\label{eq:lcwl}
	W(\xi^-,0) &\equiv P \exp\bigg(-ig \int_0^{\xi^-}d\eta^- A^+(\eta^-) \bigg) \,.
	\end{align}
	The light-cone correlator is renormalized in the $\overline{\rm MS}$ scheme at scale $\mu$. The PDF defined in \eq{lcpdf} is gauge invariant and does not depend on the hadron momentum $P^z$ or $P^+$.	In the light-cone gauge $A^+=0$, it can be expanded in terms of free-field operators, giving a clear interpretation of parton number density in the longitudinal momentum space,
	\beq \label{eq:parton}
		q_i(x)\sim \int dk^+d^2 k_\perp \delta(k^+-xP^+) \langle P| \hat{n}(k^+,\vec{k}_\perp)|P\rangle\,,
	\eeq
	where $\hat{n}(k^+,\vec{k}_\perp)$ is a quark number density operator.
	
	Since the PDF is a hadron matrix element, it is natural to calculate it from lattice gauge theory, the only practical approach to solve nonperturbative QCD so far. However, it has been extremely difficult to compute the PDFs directly from the Euclidean lattice due to the real-time dependence of the light-cone correlator. It is genuinely difficult to analytically continue the imaginary-time results from lattice to real-time to obtain the Minkowskian matrix elements. Neither can one define a nonlocal lightlike separation $z^2=0$ in the Euclidean space, as only $z^\mu=(0,0,0,0)$ satisfies this condition.
	Instead, early efforts have focused on the lattice calculation of Mellin moments of the PDF, which are given by matrix elements of local gauge-invariant twist-two operators,
	\begin{align}
		\int_{-1}^1 dx\ x^n q(x) \sim n_{\mu_0}n_{\mu_1}\cdots n_{\mu_n}\langle P | \bar{\psi}(0) \gamma^{\{\mu_0} i\overleftrightarrow{D}^{\mu_1}\cdots i\overleftrightarrow{D}^{\mu_n\}}\psi(0)|P\rangle\,,
	\end{align}
	where $n_{\mu}=(1,0,0,-1)/\sqrt{2}$, $\overleftrightarrow{D}^\mu= (\overrightarrow{\partial}^\mu - \overleftarrow{\partial}^\mu)/2 + igA^\mu$, and $\{\cdots\}$ stands for symmetric and traceless Lorentz indices. The matrix elements of these local twist-two operators can be readily calculated in lattice QCD, but the noise of calculating derivative operators increases with the number of moments. Moreover, due to the breaking of rotational symmetry on a discretized lattice, there will be power-divergent mixing between higher- and lower-dimensional operators which are under the same representation of the hypercubic $H(4)$ group, which makes it impossible to calculate higher moments without a challenging subtraction of the power divergences. Therefore, only the lowest three moments of the PDFs have been calculated in lattice QCD~\cite{Martinelli:1987zd,Martinelli:1988xs,Detmold:2001dv,Detmold:2002nf,Dolgov:2002zm}, as one can choose particular Lorentz indices to avoid such power-divergent operator mixing.
	
	Not long ago, a breakthrough was made by Ji with the proposal of the large-momentum effective theory (LaMET)~\cite{Ji:2013dva,Ji:2014gla} to calculate light-cone parton physics from lattice QCD. In LaMET, one starts from certain Euclidean observables in a large-momentum hadron state, and then match their matrix elements onto the corresponding light-cone parton observables through a factorization formula that can be systematically quantified. The quasi observable for the PDF is the so-called ``quasi-PDF" which is defined from an equal-time correlator that can be directly calculated on lattice, thus allowing for an extraction of the $x$-dependence of the PDFs from lattice QCD. Ever since it was proposed, LaMET has been actively applied to the lattice calculations of gluon helicity contribution to the proton spin~\cite{Yang:2016plb}, and, especially, the unpolarized, helicity and tranvsersity isovector quark PDFs in the proton~\cite{Lin:2014zya,Alexandrou:2015rja,Chen:2016utp,Alexandrou:2016jqi,Alexandrou:2018pbm,Alexandrou:2018eet,Lin:2017ani,Chen:2018xof,Lin:2018pvv,Liu:2018hxv} and pion~\cite{Chen:2018fwa,Karthik:2018wmj,Shugert:2018pwi,Petreczky:2018jqc}, as well as meson distribution amplitudes~\cite{Zhang:2017bzy,Chen:2017gck}. Notably, the most recent lattice calculation of the unpolarized and helicity isovector quark PDF at physical pion mass~\cite{Alexandrou:2018pbm,Chen:2018xof,Alexandrou:2018eet,Lin:2018pvv,Liu:2018hxv} and large proton momentum~\cite{Chen:2018xof,Lin:2018pvv,Liu:2018hxv} have shown remarkable agreement with the global analysis in the moderate-$x$ region, and the transversity isovector quark PDF has reached a precision that is much better than the most up-to-date fit of experimental data~\cite{Alexandrou:2018eet,Liu:2018hxv}.
	
	Meanwhile, it is worthwhile to mention that other approaches have also been proposed to calculate the PDFs from lattice QCD in recent years. To calculate higher moments of the PDFs, a method was invented to restore the rotational symmetry in the continuum limit that helps avoid the power-divergent mixings~\cite{Davoudi:2012ya}, which allows the calculation of higher moments of the PDFs. Another approach uses the operator product expansion (OPE) of the Compton amplitude from a heavy-to-light current-current correlator to extract the higher moments, where the heavy quark mass acts as the expansion parameter~\cite{Detmold:2005gg}. Recent progress with this approach has been reported in Ref.~\refcite{Detmold:2018kwu}. A direct OPE of the Compton amplitude has also been used to extract structure functions in Ref.~\refcite{Chambers:2017dov}. Besides, there are also approaches based on the factorization of Euclidean correlators in the coordinate space~\cite{Braun:2007wv,Bali:2018spj}, or more general ``lattice cross sections" which are essentially matrix elements of operators that are calculable on Euclidean lattices~\cite{Ma:2014jla,Ma:2017pxb}. Especially, the recently proposed ``pseudo-PDF" approach starts from the same correlator that defines the quasi-PDF and is based on a factorization formula in coordinate space~\cite{Radyushkin:2017cyf,Orginos:2017kos,Radyushkin:2017lvu,Karpie:2018zaz}. This factorization formla is equivalent to that for the quasi-PDF in momentum space~\cite{Ji:2017rah,Izubuchi:2018srq}, and provides an alternate way to extract the PDF from lattice matrix elements of the equal-time correlator. Finally, there is a proposal that aims at direct calculation of the physical hadronic tensor from lattice QCD to extract the structure functions or PDFs~\cite{Liu:1993cv,Liu:1998um,Liu:1999ak,Liu:2016djw,Liang:2017mye}. For a complete review of all the proposals, see Ref.~\refcite{Monahan:2018euv}.\\
	
	In this paper, we will elaborate on the systematic procedure of calculating the PDFs with the LaMET approach~\footnote{Note that a comprehensive review of the lattice calculation of PDFs with LaMET has also appeared in a recent publication~\refcite{Cichy:2018mum}.}. In \sec{theory}, we will introduce the general formalism of LaMET; In \sec{ren}, we will discuss the renormalizability of the quasi-PDF and provide a rigorous derivation of its factorization formula; In \sec{match}, we lay out a nonperturbative renormalization program for the quasi-PDF in lattice QCD, and calculate the one-loop matching kernel to extract the light-cone PDF; In \sec{pwr}, we will discuss hadron mass and higher-twist corrections which are suppressed by powers of the large momentum; We will present several recent lattice results of the isovector quark PDFs in \sec{lattice}, discuss recent developments in the calculations of gluon PDF, the generalized parton distributions (GPDs) and transvere momentum dependent (TMD) PDFs in \sec{other}, and conclude in \sec{concl}.
	
	\section{General formalism}
	\label{sec:theory}
	
	In modern language, parton physics is formulated on the light-cone ($\xi^+=0$), which is invariant under a Lorentz boost in the $z$-axis. In light-cone quantization with $A^+=0$, all partonic observables have clear particle number density interpretations, as shown in \eq{parton}. 
	Compared to the simplicity of the light-cone formalism, it is more intuitive to boost the equal-time ($x^0=0$) picture of the hadron to the infinite momentum frame (IMF), but the latter is not often used to describe parton physics in literature.
	
	In the equal-time picture, one can also define a longitudinal momentum distribution of the quark or gluon, or the quasi-PDF~\cite{Ji:2013dva},
	\begin{align}\label{eq:qpdf}
		\tilde{q}_i(x,P^z,\Lambda) \equiv \!\int\!\! \frac{dz}{4\pi} \, e^{ixP^z z} 
		\! \big\langle P \big| \bar{\psi_i} (z) \Gamma 
		W(z,0)
		\psi_i(0)\big|P\big\rangle \,,
	\end{align}
	where $\Gamma=\gamma^t$ or $\gamma^z$, $\Lambda$ is the ultraviolet (UV) momentum cutoff, and the spacelike Wilson line is
	\begin{align}\label{eq:qwl}
		W(z,0) &= P \exp\bigg(-ig \int_0^{z}dz' A^z(z') \bigg) \,.
	\end{align}
	 When the hadron is moving at finite momentum, relativistic quantum fluctuations allow the quarks and gluons to escape the hadron in the forward and backward directions, thus extending their momentum fraction $x$ to $[-\infty,\infty]$. However, unlike the PDF, the quasi-PDF is not frame independent. Owing to gauge interactions, a quark or gluon will be transformed into a quark or gluon plus an infinite number of quarks and gluons under a Lorentz boost in the longitudinal direction, so this distribution depends dynamically on the hadron momentum $P^z$. Nevertheless, in the IMF limit of $P^z\gg \Lambda$, all the vacuum pair productions are suppressed by the infinite hadron momentum~\cite{Weinberg:1966jm}, and thus quarks and gluons can only move forward with $x$ restricted to $[0,1]$. As a result, the frame-dependent quasi-PDF reduces to the frame-independent PDF in this limit.
	
	This picture of the hadron under an infinite Lorentz boost has also been used to justify the physical meaning~\cite{Ji:2012gc,Ji:2013fga,Hatta:2013gta,Ji:2014lra} of the naive sum rule of the proton spin~\cite{Jaffe:1989jz}, as frame-dependent quasi- gluon spin and quark/gluon orbital angular momentum operators all converge into the boost-invariant partonic spin and orbital angular momenta in the IMF limit.
	
	However, in a theory with UV cutoff $\Lambda$, the physical matrix elements can only be evaluated at $P^z< \Lambda$. Especially, on a lattice with spacing $a$, the hadron matrix elements can only be calculated with momentum $P^z\ll a^{-1}$, and the continuum limit $a\to0$ must be taken first to extract the physical results. In quantum field theory, the $P^z\to\infty$ and $\Lambda\to\infty$ limits are not exchangeable, and a direct extrapolation of the quasi-PDF $\tilde{q}_i(x,P^z,\Lambda)$ to $P^z\to\infty$ limit does not exist due to its dynamical dependence on $P^z$. Fortunately, when $P^z$ is much larger than the soft scales $\Lambda_{\rm QCD}$ and hadron mass $M$, changing the limits of $P^z\to\infty$ and $\Lambda\to\infty$ does not affect the contribution from the IR degrees of freedom at the scale of $\Lambda_{\rm QCD}$ and hadron mass $M$. Therefore, in the window of $\Lambda \gg P^z \gg \{\Lambda_{\rm QCD},M\}$, the difference between the quasi-PDF $\tilde{q}_i(x,P^z,\Lambda)$ and the PDF $q_i(x,\mu)$ is only in the UV part, which can be calculated analytically in perturbative QCD.
		
	The exact relation between the quasi-PDF and PDF is given by a factorization formula in LaMET~\cite{Ji:2013dva,Ji:2014gla}, which is a systematic expansion in $1/P^z$,
	\beq \label{eq:lamet}
		\tilde{q}_i(x,P^z,\Lambda) = C_{ij}(x,P^z,\Lambda,\mu) \otimes q_j(x,\mu) + O\left({\Lambda_{\rm QCD}^2\over P_z^2},{M^2\over P_z^2}\right)\,,
	\eeq
	where the second term includes all the power corrections, and the leading power contribution to the quasi-PDF can be factorized into the convolution of a perturbative matching coefficient $C$ and the PDF. $C$ only depends on the hard scales $P^z$, $\Lambda$, and $\mu$. The convolutional form can be derived rigorously from perturbative QCD, as will be shown in Sec.~\ref{sec:ren}.

	In principle, $C$ can absorb all the UV divergences in the bare matrix element of the quasi-PDF~\cite{Xiong:2013bka}, so that the renormalization is included in the matching procedure. But in practice, to extract the light-cone PDF $q_i(x,\mu)$ from the quasi-PDF $\tilde{q}_i(x,P^z,\Lambda=a^{-1})$ simulated on lattice, one usually needs to first renormalize the latter in order to take the continuum limit. The renormalization can be done in lattice perturbation theory~\cite{Capitani:2002mp}, as has been studied in Refs.~\cite{Ishikawa:2016znu,Xiong:2017jtn,Constantinou:2017sej,Ishikawa:2018wun}, but its convergence cannot be guaranteed due to the linear power divergence in the quasi-PDF~\cite{Xiong:2013bka}. Therefore, a nonperturbative renormalization scheme is favored for the lattice calculation, which will be discussed in detail in Sec.~\ref{sec:match}.
	
	\section{Renormalization and factorization}
	\label{sec:ren}
	
	As has been discussed in the previous section, it is desirable to renormalize the quasi-PDF nonperturbatively on lattice in order to take the continuum limit. First of all, one has to prove that the quasi-PDF, or the nonlocal quark bilinear operator $\tilde{O}_\Gamma(z) = \bar{\psi}(z)\Gamma W(z,0)\psi(0)$ from which the quasi-PDF is defined, is renormalizable.
	
	It is known from the first one-loop calculation~\cite{Xiong:2013bka} that the quasi-PDF includes linear power divergence that orginiates from the self-energy of the Wilson line in $\tilde{O}_\Gamma(z)$. Later, a two-loop renormalization of the quasi-PDF in the momentum space was carried out in the $\overline{\rm MS}$ scheme~\cite{Ji:2015jwa}, where the power divergence vanishes and all logarithmic UV divergences can be factored into the renormalization constant. Since then, the study has been focused on the all-order proof of the multiplicative renormalizability of the quasi-PDF.
	
	The renormalization of Wilson lines have been well studied in literature. Using the path integral formalism where an open smooth Wilson line can be replaced by the two-point function of an auxiliary field, it was elegantly proved that its renormalization is multiplicative in the coordinate space~\cite{Craigie:1980qs,Dorn:1986dt}. The proof concludes that the Wilson line $W(z,0)$ of our interest can be renormalized as
	\beq \label{eq:wlren}
		W(z,0) = Z\ e^{-\delta m|z|} W^R(z,0)\,,	
	\eeq
	where $\delta m$ is of mass dimension and includes all the linear power divergence if there is a gauge-invariant UV cutoff regulator, and $Z$ captures all the logarithmic divergences that does not depend on the coordinate $z$. The fields and couplings in the Wilson line are renormalized ones on both the l.h.s. and r.h.s. of \eq{wlren}. This relation echoes the renormalizability of closed Wilson loops that has also been proved using Ward identities~\cite{Dotsenko:1979wb}.
	
	The renormalization relation in \eq{wlren} was postulated to be applicable to nonlocal quark bilinear operators that include the same Wilson line~\cite{Musch:2010ka,Ishikawa:2016znu}, i.e.
	\beq \label{eq:ren}
	\tilde{O}_\Gamma(z) = Z_{\psi,Z}\ e^{-\delta m|z|} \tilde{O}_\Gamma^R(z)\,,	
	\eeq
	and was verified at one loop in the transverse-momentum cutoff regularization scheme~\cite{Chen:2016fxx}. Here $\delta m$ is the same as that in the Wilson line self-energy, and $Z_{\psi,Z}$ includes additional logarithmic divergences at the end points as well as the quark wavefunction renormalization. This relation was later proved independently using the auxiliary field method~\cite{Ji:2017oey,Green:2017xeu} and a full diagrammatic analysis~\cite{Ishikawa:2017faj}. Moreover, the renormalization is diagonal in quark flavors and has no mixing with the gluon quasi-PDF, that is,
	\beq \label{eq:renmix}
	\bar{\psi_i} (z) \Gamma W(z,0)\psi_i(0) = Z_{\psi,Z}\ e^{-\delta m|z|} \left[\bar{\psi_i} (z) \Gamma W(z,0)\psi_i(0)\right]^R\,.
	\eeq
	
	The above proof has also been generalized to the gluon case, as the equal-time correlator used to define the gluon quasi-PDF can also be multiplicatively renormalized in the coordinate space~\cite{Zhang:2018diq,Li:2018tpe} up to a contact term~\cite{Zhang:2018diq}.\\
	
	Based on the above results, we can study the OPE of the renormalized nonlocal quark bilinear $\tilde{O}_\Gamma(z,\mu)$ in the $\overline{\rm MS}$ scheme to derive the factorization formula for the quasi-PDF. For simplicity, we take $\Gamma=\gamma^z$, and the conclusion also applies to the $\Gamma=\gamma^t$ case. 
	
	In the limit of $|z|\to 0$, $\tilde{O}_\Gamma(z,\mu)$ can be expanded 
	in terms of local gauge-invariant operators~\cite{Izubuchi:2018srq},
	\begin{align}\label{eq:ope-tq}
	\tilde{O}_{\gamma^z}(z,\mu) = & \sum_{n=0}^\infty \left[C^q_n ({\mu}^2 z^2)\frac{(-iz)^n}{n!} e_{\mu_1}\cdots 				e_{\mu_n}O_q^{\mu_0\mu_1\cdots\mu_n}(\mu)\right. \nn\\
	&+C^g_n ({\mu}^2 z^2)\frac{(-iz)^n}{n!} e_{\mu_1}\cdots e_{\mu_n}O_g^{\mu_0\mu_1\cdots\mu_n}(\mu) +  \text{higher-twist terms}\Big]\,,
	\end{align}
	where $e^\mu=(0,0,0,1)$, $\mu_0=z$, $C^q_n=1+O(\alpha_s)$ and $C^g_n=O(\alpha_s)$ are the Wilson coefficients, and $O_q^{\mu_0\mu_1\cdots\mu_n}(\mu)$ and $O_g^{\mu_0\mu_1\cdots\mu_n}(\mu)$ are the only allowed renormalized traceless symmetric twist-2 quark and gluon operators at leading power in the OPE,
	\begin{align}\label{eq:twist-2}
	O_q^{{\mu}_0 {\mu}_1 \ldots {\mu}_n}(\mu) = & Z_{n+1}^{qq} \bigl(
	\bar{\psi} \gamma^{({\mu}_0 } iD^{{\mu}_1} \cdots iD^{ {\mu}_n)} \psi -
	\text{trace} \bigr) \,,
	\\
	O_g^{{\mu}_0 {\mu}_1 \ldots {\mu}_n}(\mu) = & Z_{n+1}^{qg}\bigl(
	F^{(\mu_0\rho} iD^{{\mu}_1} \cdots iD^{ {\mu}_{n-1}} F_\rho^{~\mu_n)} - \text{trace} \bigr)
	.\nn
	\end{align}
	Here $Z_{n+1}^{ij}=Z_{n+1}^{ij}(\epsilon)$ are multiplicative $\overline{\rm MS}$ renormalization factors and $(\mu_0 \cdots \mu_n)$ stands for the symmetrization of these Lorentz indices.
	
	The above OPE is valid for the operator itself, where we implicitly restrict the expansion to the local operators whose forward matrix elements do not vanish. Moreover, we only consider the iso-vector case $i=u-d$, so the mixing with the gluon operators is absent. When $O_q^{\mu_0\mu_1\cdots\mu_n}$ is evaluated in the hadron state $|P\rangle$,
	\beq \label{eq:def-moments}
	\langle P | O_q^{{\mu}_0 {\mu}_1 \cdots {\mu}_n} | P\rangle =  2a_{n + 1}(\mu)\left(P^{{\mu}_0} P^{{\mu}_1} \ldots P^{{\mu}_n} - \text{trace}\right)\,,
	\eeq
	where $a_{n+1}(\mu)$ is the $(n+1)$-th moment of the PDF,
	\beq \label{eq:moment}
	a_{n + 1} \left(\mu\right)=\int_{-1}^1 dx\,x^n q \left(x,\mu\right)\,.
	\eeq
	When contracted with $e_{\mu_1}\cdots e_{\mu_n}$, the trace contribution in \eq{def-moments} is exactly the mass corrections of $O(M^2/P_z^2)$ in \eq{lamet}, whose explicit expression of the trace term have been derived in Ref.~\cite{Nachtmann:1973mr,Georgi:1976ve,Chen:2016utp} and will be discussed in \sec{pwr}.
	
	The Wilson coefficients $C^q_n(\mu^2 z^2)$ in the OPE of $\tilde{O}_{\gamma^z}(z)$ can be calculated in perturbation theory, and the one-loop results are~\cite{Izubuchi:2018srq}
	\begin{align} \label{eq:wilson}
		C^q_n(\mu^2 z^2) =& 1+{\alpha_sC_F\over 2\pi}\left[\left({3+2n\over 2+3n+n^2}+2H_n\right)\ln{\mu^2z^2e^{2\gamma_E}\over4}\right.\nn\\
		&\left.+{7+2n\over 2+3n+n^2}+2(1-H_n)H_n -2H_n^{(2)}\right]\,,	
	\end{align}
	where $C_F=4/3$, and the Harmonic numbers are $H_n=\sum_{i=1}^n 1/i$ and $H_n^{(2)}=\sum_{i=1}^n 1/i^2$.
	
	Based on Eqs.~(\ref{eq:def-moments}-\ref{eq:wilson}), we can obtain the leading-twist approximation of the hadron matrix element of the nonlocal quark bilinear $\tilde{O}_{\gamma^z}(z,\mu)$ as
	\begin{align} \label{eq:ldtwist}
	\frac{\langle P | \tilde{O}_{\gamma^z}(z,\mu) | P \rangle}{2P^z} 
	=&  \sum_{n} C^q_n (\mu^2z^2)
	\frac{(i zP^z)^n}{n!}  \int_{-1}^1 dy\,y^n q \left( y, \mu\right) + O\left(z^2\Lambda_{\rm QCD}^2,{M^2\over P_z^2}\right)\,.
	\end{align}
	Note that in the above approximation we have used $|z|\ll \Lambda_{\rm QCD}^{-1}$ and implied $P^z\gg M$, but the latter limit is not compulsory as we can keep the kinematic trace terms in the leading-twist contribution, and it is known how to absorb them into the quasi-PDF~\cite{Chen:2016utp}. In the large momentum limit $P^z\gg \{\Lambda_{\rm QCD},M\}$, we will have $P^z\sim P^0$, so \eq{ldtwist} still stands for $\tilde{O}_{\gamma^0}(z,\mu)$ except that the Wilson coefficients $C^q_n$ are modified by a finite constant~\cite{Izubuchi:2018srq}.
	
	After Fourier transforming to the momentum space, we obtain the quasi-PDF in the $\overline{\rm MS}$ scheme and derive the factorization formula for it~\cite{Izubuchi:2018srq},
	\begin{align} \label{eq:quasi}
		\tilde{q} \Bigl(  x, P^z,\mu\Bigr)
		\equiv &\int \frac{d z}{4 \pi P^z}\: e^{ix P^zz}\:  \langle P | \tilde{O}_{\gamma^z}(z,\mu) | P \rangle \nn\\
		= &  \int_{-1}^1 {dy\over |y|} \biggl[\int \frac{d (yP^zz)}{2 \pi} e^{i
			\frac{x}{y}\cdot yP^zz}  \sum_{n=0} C^q_n \Bigl( \frac{{\mu}^2(yP^zz)^2 }{(yP^z)^2}\Bigr)\,  \frac{(-i yP^zz)^n}{n!}  \biggr] q \left( y,\mu \right)\nn\\
		& +O\left({\Lambda_{\rm QCD}^2\over x^2P_z^2},{M^2\over P_z^2}\right) \,.
	\end{align}
	It is evident that the matching kernel is a function of $x/y$ and $\mu/(|y|P^z)$, so we define it as
	\begin{align}   \label{eq:kernel}
		C \left( {x\over y}, \frac{\mu}{|y|P^z} \right)
		\equiv \int \frac{d (yP^zz)}{2 \pi} e^{i
			\frac{x}{y}\cdot P^zz}  \sum_{n=0} C^q_n \Bigl( \frac{{\mu}^2(yP^zz)^2 }{(yP^z)^2}\Bigr)\,  \frac{(-i yP^zz)^n}{n!}
		\,,
	\end{align}
	and rewrite \eq{lamet} as
	\begin{align} \label{eq:msfac}
		\tilde{q} \left( x, P^z,\mu\right)
		= \int_{-1}^1 \frac{dy}{|y|}\: C \Bigl(
		\frac{x}{y}, \frac{\mu}{|y|P^z} \Bigr)\: q \left( y,\mu\right)+O\left({\Lambda_{\rm QCD}^2\over x^2P_z^2},{M^2\over P_z^2}\right)
		\,.
	\end{align}
	Compared to the same formula derived before~\cite{Izubuchi:2018srq}, the higher-twist contributions are suppressed by powers of $xP^z$ instead of $P^z$, because it is $xP^z$ that is the Fourier conjugate to $z$. Therefore, the small-$x$ quasi-PDF includes larger higher-twist corrections. Note that in a recent work on the renormalon ambiguity in the factorization formula in \eq{msfac}~\cite{Braun:2018brg} for $\overline{\rm MS}$ quasi-PDF, it was concluded that the power corrections include a term that scales as $\Lambda_{\rm QCD}^2/[x^2(1-x)P_z^2]$, which indicates that power corrections at $x\to1$ is also important.
	
	The factorization formula in \eq{msfac} is derived for the quasi-PDF in the $\overline{\rm MS}$ scheme, but it can be easily generalized to the other renormalization schemes~\cite{Izubuchi:2018srq}. Since the quasi-PDF is multiplicatively renormalizable in the coordinate space, we can convert between different schemes perturbatively, which in effect proves the factorization formula for the quasi-PDF in other schemes than $\overline{\rm MS}$.
	
	\section{Nonperturbative renormalization on lattice and matching}
	\label{sec:match}
	
	As has been explained in \sec{theory}, it is desirable to perform a nonperturbative renormalization of the quasi-PDF on lattice first so that we can extract its continuum limit. Several different nonperturbative schemes have been proposed in literature.
	
	One of the choices is to determine $\delta m$ in Eqs.~(\ref{eq:wlren}) and (\ref{eq:ren}) nonperturbatively from the static quark-antiquark potential~\cite{Musch:2010ka,Ishikawa:2016znu,Zhang:2017bzy,Green:2017xeu}. $\delta m$ includes a linearly divergent term $\propto 1/a$ which will cancel such divergence in the nonlocal quark bilinear $\tilde{O}_\Gamma(z)$ on lattice. After the nonperturbative subtraction of linear power divergence from the quasi-PDF, one can then renormalize the logarithmic divergences using either lattice perturbation theory~\cite{Ishikawa:2016znu,Xiong:2017jtn,Constantinou:2017sej} or other nonperturbative schemes for local composite operators~\cite{Green:2017xeu}. However, to match the renormalized quasi-PDF in this mixed scheme, one has to calculate $\delta m$ perturbatively in the corresponding scheme, which is currently not clear and might have to be done in lattice perturbation theory.
	
	A more direct nonperturbative renormalization can be performed in the regulator-independent momentum subtraction (RI/MOM) scheme~\cite{Martinelli:1994ty}, which has been widely used for the lattice renormalization of local composite quark operators that are free from power-divergent mixings. Since the nonlocal quark bilinear $\tilde{O}_\Gamma(z)$ is of the lowest mass dimension of its kind, it cannot mix with other operators in a power-divergent way. Given that it is multiplicatively renormalizable in coordinate space, we can also implement the RI/MOM scheme for the quasi-PDF on lattice~\cite{Constantinou:2017sej,Stewart:2017tvs}.
	
	Before we proceed with the RI/MOM renormalization, it should be noted that due to the breaking of chiral symmetry in Wilson-type fermion actions, there is additional operator mixing in $\tilde{O}_\Gamma(z)$~\cite{Constantinou:2017sej,Green:2017xeu,Chen:2017mie}. For $\Gamma=\gamma^z$, the operator can mix with the scalar case with $\Gamma=\cal I$ at $O(a^0)$. For $\Gamma=\gamma^t$, there is no mixing at $O(a^0)$. To reduce the systematic uncertainties from lattice renormalization, it is preferred to choose $\Gamma=\gamma^t$ and $\Gamma=\gamma^5\gamma^z$ for the unpolarized and polarized quark PDF respectively.
	
	For each $z$, the RI/MOM renormalization factor $Z$ is calculated nonperturbatively on lattice by imposing the condition that the renormalized quasi-PDF in an off-shell quark state is equal to the tree-level value at a perturbative subtraction point,
	\beq \label{eq:Z}
		\left.Z^{-1}(z,p^R_z,\mu_R,a^{-1})\langle p|\tilde{O}_{\gamma^t}(z)|p\rangle \right|_{\tiny\begin{matrix}p^2=-\mu_R^2 \\ \!\!\!\!p_z=p^R_z\end{matrix}}= \langle p|O_{\gamma^t}(z)|p\rangle_{\rm tree}\,,
	\eeq
	The bare matrix element $\langle p|\tilde{O}_{\gamma^t}(z)|p\rangle$ is defined from the amputated Green's function $\Lambda_{\gamma^t}(p,z)$ of the nonlocal quark bilinear $\tilde{O}_{\gamma^t}(z)$. Since the Green's function is not gauge invariant, one needs to fix the gauge condition on lattice, which is usually chosen to be the Landau gauge. $\Lambda_{\gamma^t}(p,z)$ is calculated nonperturbatively on lattice with a projection operator ${\cal P}$ for the Dirac matrix,
	\begin{equation}
	\sum_s\langle p,s|O_{\gamma^t}(z)|p,s\rangle = \mbox{Tr}\left[ \Lambda_{\gamma^t}(z,p){\cal P}\right]\,.
	\end{equation}
	Since $\tilde{O}_{\gamma^t}(z)$ is not boost invariant in the $z$ direction, its matrix element will depend not only on the four-momentum squared $p^2$, but also on the momentum component $p^z$, so we have to fix both scales at the subtraction point $p^2=-\mu_R^2$, $p_z=p^R_z$. On lattice, the external momentum $p_E^\mu$ is Euclidean, so the subtraction point $p_E^2=\mu_R^2$. As a result, the renormalization factor $Z(z,p^R_z,\mu_R,a^{-1})$ also must depend on $\mu_R$ and $p_z^R$. To work in the pertubrative region and control the lattice discretization error of order $O(a^2\mu_R^2, a^2(p_z^R)^2)$, one must work in the window $\Lambda_{\rm QCD}\ll \mu_R\ll a^{-1}$, $p_z^R \ll a^{-1}$, which is attainable if the lattice spacing is small enough.
	
	According to the $H(4)$ symmetry on lattice, the amputated Green's function $\Lambda_{\gamma^t}(p,z)$ is not only proportional to the tree-level matrix element $\gamma^t$, but also includes two other independent Lorentz structures~\cite{Liu:2018uuj}:
	\begin{equation}\label{eq:ME_decomposition}
	\Lambda_{\gamma^t}(p,z)=\tilde{F}_t(p,z) \gamma^t  + \tilde{F}_{z}(p,z)\frac{p_t\gamma^z}{p_z} +\tilde{F}_p(p,z)\frac{p_t\slashed{p}}{p^2}\,,
	\end{equation}
	where $\tilde{F}_i$'s are form factors that are invariant under the hyper cubic group $H(4)$. Therefore, $Z(z,p^R_z,\mu_R,a^{-1})$ also depends on the projection operator ${\cal P}$. $\tilde{F}_t$ includes all the UV divergences in $\Lambda_{\gamma^t}(p,z)$, while $\tilde{F}_z$ and $\tilde{F}_p$ are UV finite.	
	We can choose ${\cal P}$ to only sort out $\tilde{F}_t$~\cite{Stewart:2017tvs}, which is named as the minimal projection~\cite{Liu:2018uuj}. On the other hand, we can also choose ${\cal P} = \slashed p/(4p^t)$~\cite{Stewart:2017tvs}, which is named as the $\slashed p$ projection~\cite{Liu:2018uuj}. The renormalization factor $Z$ with the minimal and $\slashed p$ projections are
	\begin{align}\label{eq:def_renorm}
		Z_{mp}(z,p^R_z,\mu_R,a^{-1})\equiv &\tilde{F}_t(p,z)\Big|_{\tiny\begin{matrix}p^2_E=\mu_R^2 \\ \!\!\!\!p_z=p^R_z\end{matrix}},\\
		Z_{\slashed{p}}(z,p^R_z,\mu_R,a^{-1})\equiv &\Big[\tilde{F}_t(p,z)+\tilde{F}_z(p,z)+\tilde{F}_p(p,z)\Big]\bigg|_{\tiny\begin{matrix}p^2_E=\mu_R^2 \\ \!\!\!\!p_z=p^R_z\end{matrix}}.
	\end{align}
	
	Then, the bare hadron matrix element of the $\tilde{O}_{\gamma^t}(z)$
	\begin{align}
	\tilde{h}(z,P_z,a^{-1}) = {1\over 2P^0} \langle P |\tilde{O}_{\gamma^t}(z)| P \rangle
	\end{align}
	is renormalized in coordinate space as
	\beq \label{eq:rimomh}
	\tilde{h}_R(z,P_z, p^R_z,\mu_R)=\lim_{a\to0}Z^{-1}(z,p^R_z,a^{-1},\mu_{\tiny R})\tilde{h}(z,P_z,a^{-1}) \,,
	\eeq
	where $\tilde{h}_R(z,P_z, p^R_z,\mu_R)$ is the renormalized matrix element. At finite lattice spacing $a$, $\tilde{h}_R(z,P_z, p^R_z,\mu_R)$ could still have discretization errors which is polynomial in $a$, so one is expected to perform calculation at different spacings and extrapolate to the continuum limit.\\
	
	The next step is to match the renormalized quasi-PDF to the $\overline{\rm MS}$ PDF. Since the renormalized matrix element is indepenent of the UV regulator, we should obtain the same result even in dimensional regularization under the same renormalization scheme, i.e.,
	\beq
		\lim_{a\to0}Z^{-1}(z,p^R_z,\mu_R,a^{-1})\tilde{h}(z,P_z,a^{-1}) = \lim_{\epsilon\to0}Z^{-1}(z,p^R_z,\mu_R,\mu,\epsilon)\tilde{h}(z,P_z,\mu,\epsilon)\,,
	\eeq
	which allows us to compute the matching coefficients in continuum theory. Here $\mu$ is the scale introduced in dimensional regularization.
	
	Two strategies have been developed for the matching procedure in literature~\cite{Constantinou:2017sej,Stewart:2017tvs}: One strategy is to first convert the RI/MOM matrix element $\tilde{h}_R(z,P_z, p^R_z,\mu_R)$ into the $\overline{\rm MS}$ scheme in coordinate space~\cite{Constantinou:2017sej}, and then Fourier transform to momentum space to obtain the $\overline{\rm MS}$ quasi-PDF, and eventually match the latter to the $\overline{\rm MS}$ PDF~\cite{Izubuchi:2018srq}. The other strategy is to first Fourier transform $\tilde{h}_R(z,P_z, p^R_z,\mu_R)$ into momentum space to obtain the RI/MOM quasi-PDF, and then match the latter directly to the $\overline{\rm MS}$ PDF~\cite{Stewart:2017tvs}. Both strategies are equivalent in principle.
	
	For the first strategy, the conversion factor $Z^{\rm RI/MOM}_{\overline{\rm MS}}$ between the RI/MOM and $\overline{\rm MS}$ scheme is the difference between $Z(z,p^R_z,\mu_R,\mu,\epsilon)$ ~\cite{Constantinou:2017sej} and the $\overline{\rm MS}$ renormalization constant $Z_{\overline{\rm MS}}(\epsilon)$,
	\beq
		Z^{\rm RI/MOM}_{\overline{\rm MS}}(z,p_z^R,\mu_R,\mu) = \frac{Z(z,p^R_z,\mu_R,\mu,\epsilon)}{Z_{\overline{\rm MS}}(\epsilon)} = \frac{Z(z,p^R_z,\mu_R,\mu,\epsilon)}{\displaystyle 1+{\alpha_sC_F\over 2\pi}{3\over 2}\left[{1\over\epsilon}-\gamma_E+\ln(4\pi)\right]+O(\alpha_s^2)} \,.\nn\\
	\eeq
	In the $\overline{\rm MS}$ quasi-PDF, the IR divergences are regulated with dimensional regularization as the quarks are massless. In Ref.~\cite{Spanoudes:2018zya}, renormalization of the quasi-PDF have also been provided for finite quark masses.
	
	 The matching coefficient for the quasi-PDF in the $\overline{\rm MS}$ scheme is~\cite{Izubuchi:2018srq}
	 \begin{align} \label{eq:quasi-matching}
	 &C^{\overline{\rm MS}}\left(\xi, {\mu\over |y| P^z}\right)-\delta\left(1-\xi\right)\nn\\
	 = & {\alpha_sC_F\over 2\pi}\left\{
	 \begin{array}{ll}
	 \displaystyle \left({1+\xi^2\over 1-\xi}\ln {\xi\over \xi-1} + 1 + {3\over 2\xi}\right)^{[1,\infty]}_{+(1)}- {3\over 2\xi}
	 &\, \xi>1
	 \nn\\[10pt]
	 \displaystyle \left({1+\xi^2\over 1-\xi}\left[
	 - \ln{\mu^2\over y^2P_z^2} + \ln\big(4\xi(1-\xi)\big)\right] - {\xi(1+\xi)\over 1-\xi}\right)^{[0,1]}_{+(1)}
	 &\, 0<\xi<1
	 \nn\\[10pt]
	 \displaystyle  \left(-{1+\xi^2\over 1-\xi}\ln {-\xi\over 1-\xi} - 1 + {3\over 2(1-\xi)}\right)^{[-\infty,0]}_{+(1)} - {3\over 2(1-\xi)}\quad
	 &\, \xi<0
	 \end{array}\right.\nn\\[5pt]
	 & + {\alpha_sC_F\over 2\pi}\delta(1-\xi) \left( {3\over2}\ln{\mu^2\over 4y^2P_z^2} + {5\over2}\right)\,,
	 \end{align}
	where $\xi=x/y$, and the plus functions are defined in a general way as
	\beq
		\int_D dx\ [h(x)]_{+(x_0)}^D g(x) = \int_D dx\ h(x) [g(x) - g(x_0)]\,,
	\eeq
	where $D$ is the domain where the function $h(x)$ is defined.
	
	In the limit of $|z|\to0$, the conversion factor is logarithmically divergent, 
	\beq
		\lim_{|z|\to0}Z^{\rm RI/MOM}_{\overline{\rm MS}}(z,p_z^R,\mu_R,\mu) =1 + {\alpha_sC_F\over 2\pi}\left[{3\over 2}\ln\left({\mu^2 z^2 e^{2\gamma_E}\over 4}\right) +{5\over2}\right]+O(\alpha_s^2)\,,
	\eeq
	and $C^{\overline{\rm MS}}$ does not satisfy vector current conservation, which could potentially make numerical implementations intriciate. Note that this limit is determined by the short distance property of the nonlocal Wilson line operator, so it is independent of the external state parameters $p_z^R$ or $\mu_R$. Therefore, it was proposed that one changes the $\overline{\rm MS}$ scheme to the ``ratio scheme"~\cite{Izubuchi:2018srq}, and the conversion factor is perturbatively modified as
	\beq
		Z^{\rm RI/MOM}_{\rm ratio}(z,p_z^R,\mu_R,\mu)= \frac{Z^{\rm RI/MOM}_{\overline{\rm MS}}(z,p_z^R,\mu_R,\mu)}{\displaystyle 1+{\alpha_sC_F\over 2\pi}\left[{3\over 2}\ln(\mu^2 z^2 e^{2\gamma_E}/4) +{5\over2}\right]+O(\alpha_s^2)}\,,
	\eeq
	with
	\beq
		\lim_{|z|\to0}Z^{\rm RI/MOM}_{\rm ratio}(z,p_z^R,\mu_R,\mu) =1 +O(\alpha_s^2)\,,
	\eeq
	and the corresponding matching coefficient 
	\begin{align} \label{eq:quasi-matching-alt}
	&C^{\rm ratio\!}\left(\xi, {\mu\over |y| P^z}\right)-\delta\left(1-\xi\right)\nn\\
	=&{\alpha_sC_F\over 2\pi}\left\{
	\begin{array}{ll}
	\displaystyle \left({1+\xi^2\over 1-\xi}\ln {\xi\over \xi-1} + 1 - {3\over 2(1-\xi)}\right)^{[1,\infty]}_{+(1)}
	&\, \xi>1
	\nn\\[10pt]
	\displaystyle \left({1+\xi^2\over 1-\xi}\left[
	- \ln{\mu^2\over y^2P_z^2} + \ln\big(4\xi(1-\xi)\big)-1\right] +1+ {3\over 2(1-\xi)}\right)^{[0,1]}_{+(1)}
	&\, 0<\xi<1
	\nn\\[10pt]
	\displaystyle  \left(-{1+\xi^2\over 1-\xi}\ln {-\xi\over 1-\xi} - 1 + {3\over 2(1-\xi)}\right)^{[-\infty,0]}_{+(1)}\quad
	&\, \xi<0
	\end{array}\right.\,.\nn\\
	\end{align}
	satisfies vector current conservation~\cite{Izubuchi:2018srq}.
	
	This two-step matching procedure with $\overline{\rm MS}$ (or a different modified $\overline{\rm MS}$) as intermediate scheme has been implemented in the lattice calculations of iso-vector quark PDFs by the European Twisted Mass (ETM) Collaboration~\cite{Alexandrou:2017huk,Alexandrou:2018pbm,Alexandrou:2018eet}.\\
	
	The other strategy for matching the quasi-PDF in the RI/MOM scheme is more straightforward~\cite{Stewart:2017tvs}. If we define the Fourier transform of the conversion factor to be
	\beq
		\bar{Z}^{\rm RI/MOM}_{\overline{\rm MS}}(\eta,{\mu_R\over p_z^R},{\mu\over \mu_R})\equiv p_z^R \int {dz\over 2\pi}e^{i\eta p_z^R z}Z^{\rm RI/MOM}_{\overline{\rm MS}}(z,p_z^R,\mu_R,\mu)\,,
	\eeq
	then the matching coefficient between the RI/MOM quasi-PDF and $\overline{\rm MS}$ PDF is~\cite{Izubuchi:2018srq}
	\beq
		C^{\rm RI/MOM}\left({x\over y}, {\mu_R\over p_z^R},{\mu \over yP^z}, {yP^z\over p_z^R}\right)=\int d\eta\ \bar{Z}^{\rm RI/MOM}_{\overline{\rm MS}}\big(\eta,{\mu_R\over p_z^R},{\mu\over \mu_R}\big) C^{\overline{\rm MS}}\left({x\over y}-{\eta\over y}{p_z^R\over P^z},{\mu \over yP^z}\right)\,.\nn\\
	\eeq
	
	At one-loop order, the matcing coefficient $C^{\rm RI/MOM}$ has been computed with off-shell quark states~\cite{Stewart:2017tvs,Liu:2018uuj} for different $\Gamma$ and projection operators $\cal P$, 
	\begin{align}\label{eq:matching_coeff}
	&C^{\rm RI/MOM}\left(\xi,r,\frac{p_z}{\mu},\frac{p_z}{p_z^R}\right)\nn\\
	=&\delta(1-\xi)+\left[\tilde{q}^{(1)}_B(\xi,\rho)-q^{(1)}(\xi,p,\mu)-\tilde{q}_{CT}^{(1)}\left(\xi,r,\frac{p_z}{p_z^R}\right)\right]\Bigg|_{\cal P}+{\cal O}(\alpha_s^2)\nonumber\\
	=&\delta(1-\xi)+\left[f_0\left(\xi,\frac{p_z}{\mu}\right)-\left|\frac{p_z}{p_z^R}\right|f_{\cal P}\left(1+\frac{p_z}{p_z^R}(\xi-1),r\right)\right]_+ +{\cal O}(\alpha_s^2)\,,
	\end{align}
	where $\rho=-p^2/(p_z^R)^2$, $r=\mu_R^2/(p_z^R)^2$, $p^z=yP^z$. $\tilde{q}_B^{(1)}$ and $q^{(1)}$ are one-loop corrections to the bare quasi-PDF and $\overline{\rm MS}$ light-cone PDF, and $\tilde{q}_{CT}^{(1)}$ is the one-loop RI/MOM counter-term to the quasi-PDF in the Euclidean space (with $r>1)$. $f_0$ is the difference between $\tilde{q}_B^{(1)}$ and $q^{(1)}$ in the large momentum and on-shell limit $\rho\to0$, and is independent of $\cal P$ and the gauge choice for the bare quasi-PDF, whereas $f_{\cal P}$ comes from $\tilde{q}_{CT}^{(1)}$ and depends on $\cal P$ as well as the gauge condition.
	
	It should also be noted that the matching coefficient $C^{\rm RI/MOM}$ guarantees the vector current conservation.
	
	Without loss of generality, in the Landau gauge, the results for $\Gamma=\gamma^t$ and mimimal projection at one-loop order are~\cite{Liu:2018uuj}
	\begin{align}
		f_0\left(\xi,\frac{p_z}{\mu}\right)=\frac{\alpha_s C_F}{2\pi}\left\{
		\begin{array}{lc}
		\displaystyle \frac{1+\xi^2}{1-\xi}\ln\frac{\xi}{\xi-1}+1 &\ \xi>1\\
		\displaystyle \frac{1+\xi^2}{1-\xi}\ln\frac{4\xi(1-\xi)p_z^2}{\mu^2}-\frac{\xi(1+\xi)}{1-\xi} &\ 0<\xi<1\\
		\displaystyle -\frac{1+\xi^2}{1-\xi}\ln\frac{\xi}{\xi-1}-1 &\ \xi<0
		\end{array} \right.\,,
	\end{align}
	and 
	\begin{align}
	f_{\cal P}(\xi,r)=\widetilde{f}_t(\xi,r)=&\frac{\alpha_s C_F}{2\pi}\left\{
	\begin{array}{lc}
	\displaystyle \frac{-3r^2+13r\xi-8\xi^2-10r\xi^2+8\xi^3}{2(r-1)(\xi-1)(r-4\xi+4\xi^2)} & \\
	\displaystyle +\frac{-3r+8\xi-r\xi-4\xi^2}{2(r-1)^{3/2}(\xi-1)}\tan^{-1}\frac{\sqrt{r-1}}{2\xi-1} &\ \xi>1\\
	\displaystyle \frac{-3r+7\xi-4\xi^2}{2(r-1)(1-\xi)} &\\
	\displaystyle +\frac{3r-8\xi+r\xi+4\xi^2}{2(r-1)^{3/2}(1-\xi)}\tan^{-1}\sqrt{r-1} &\ 0<\xi<1\\
	\displaystyle -\frac{-3r^2+13r\xi-8\xi^2-10r\xi^2+8\xi^3}{2(r-1)(\xi-1)(r-4\xi+4\xi^2)}&\\
	\displaystyle -\frac{-3r+8\xi-r\xi-4\xi^2}{2(r-1)^{3/2}(\xi-1)}\tan^{-1}\frac{\sqrt{r-1}}{2\xi-1} &\ \xi<0
	\end{array} \right. .
	\end{align}
	
	More complete results can be found in Refs.~\refcite{Stewart:2017tvs,Liu:2018uuj}. This strategy has been implemented in the lattice calculations of the isovector quark PDFs in the proton and pion by the Lattice Parton Physics Project (LP$^3$) Collaboration~\cite{Chen:2017mzz,Lin:2017ani,Chen:2018xof,Chen:2018fwa,Liu:2018uuj,Lin:2018pvv,Liu:2018hxv}.\\
	
	At last, it should be noted that there is another distinct method to renormalize the quasi-PDF on lattice, which is based on a redefinition of the quasi-PDF with the gradient flow method~\cite{Monahan:2016bvm}. The smeared quasi-PDF with the gradient flow remains finite in the continuum limit, which in principle avoids the renormalization on lattice, and can be perturbatively matched on the light-cone PDF~\cite{Monahan:2017hpu}.

	\section{Power corrections}
	\label{sec:pwr}
	
	In the most recent calculations done by the LP$^3$ and ETM collaborations~\cite{Alexandrou:2018pbm,Alexandrou:2018eet,Chen:2018xof,Lin:2018pvv,Liu:2018hxv}, the largest proton momentum on lattice are 3.0 GeV and 1.4 GeV respectively. By naive power counting, the mass and higher-twist corrections can be non-negligible, and a systematic subtraction of them is necessary for precision calculations.
	
	The mass correction is purely kinematic, and is essentially the target-mass correction that is known in literature~\cite{Nachtmann:1973mr,Georgi:1976ve}. We can implement it on the quasi-PDF through the following formula~\cite{Chen:2016utp},
	\begin{align}\label{eq:mass}
		\tilde{\tilde{q}}(x) =& \sqrt{1+4c}\sum_{n=0}^\infty {f^n_- \over f^{n+1}_+} \left[(1+(-1)^n)\tilde{q}\left({f^{n+1}_+ x\over 2f_-^n}\right)+(1-(-1)^n)\tilde{q}\left(-{f^{n+1}_+ x\over 2f_-^n}\right)\right] \nn\\
		=& \sqrt{1+4c}\sum_{n=0}^\infty {(4c)^n\over f^{2n+1}_+} \left[(1+(-1)^n)\tilde{q}\left({f^{2n+1}_+ x\over 2(4c)^n}\right)+(1-(-1)^n)\tilde{q}\left(-{f^{2n+1}_+ x\over 2(4c)^n}\right)\right]\,,
	\end{align}
	where $c=M^2/(4P_z^2)$, and $f_\pm = \sqrt{1+4c}\pm1$.
	
	In practice, one truncates the above series at fixed $n$. If we truncate it at $n=0$, \eq{mass} reduces to
	\beq
		\tilde{\tilde{q}}(x) = \frac{2\sqrt{1+4c}}{\sqrt{1+4c}+1} \tilde{q}\left({(\sqrt{1+4c}+1) x\over 2}\right)\,.
	\eeq
	
	As for the corrections of order $O(\Lambda_{\rm QCD}^2/P_z^2)$, they originate from the higher-twist contributions. One can sum the higher-twist terms in the OPE in \eq{ope-tq}, and then calculate them from lattice QCD. The explicit form of the twist-4 contribution to the quasi-PDF has been given in Ref.~\refcite{Chen:2016utp},
	\beq
		\tilde{q}_{\tiny\rm twist-4}(x,P^z,\Lambda)={1\over 8\pi}\int dz\ \Gamma_0(-ix P^z z)\langle P| \tilde{O}_{\rm tr}(z) | P\rangle \,,	
	\eeq
	where $\Gamma_0$ is the incomplete Gamma function, and
	\begin{align}
		\tilde{O}_{\rm tr}(z) =& \int_0^z dz_1\ \bar{\psi}(0)\left[\gamma^\nu W(0,z_1) D_\nu W(z_1,z)\right.\nn\\
		&\left. + \int_0^{z_1} \Gamma W(0,z_2)D^\nu W(z_2,z_1) D_\nu W(z_1,z) \right]\psi(z)\,.
	\end{align}
	
	As has been discussed in \sec{ren}, it is necessary to perform a lattice renormalization of the matrix elements of the quasi-PDF, which includes the higher-twist contributions such as $\tilde{q}_{\rm twist-4}$. However, the renormalization of $\tilde{O}_{\rm tr}(z)$ is highly nontrivial, as it can mix with the lower dimensional operator $\tilde{O}_\Gamma(z)$ on lattice with a power-divergent coefficient, which would require a nontrivial subtraction for us to take the continuum limit. This is an open question that remains to be answered, and can be important to reducing the systematic uncertainties in the final result of the PDF at small-$x$.
	
	For current calculations, one can postpone the $O(\Lambda_{\rm QCD}^2/P_z^2)$ subtraction and apply the other systematic corrections for quasi-PDF at different $P^z$'s first. At the final step, one can use a simple form $A(x)+{B(x)/P_z^2}$ to extrapolate the results to the $P^z\to\infty$ limit to eliminate all the higher-twist corrections. Nevertheless, the effectiveness of this treatment should be investigated for small-$x$ distributions as the higher-twist effects in this region is enhanced by $1/x^2$.
	
	\section{Selected lattice results}
	\label{sec:lattice}
	
	With the development of nonperturbative renormalization, significant progress has been made on the lattice calculation of PDFs using the LaMET approach.
	
	The first lattice calculations at physical pion mass with nonperturbative renormalization have been independently carried out by the ETM and LP$^3$ collaborations. So far, the proton isovector unpolarized~\cite{Alexandrou:2018pbm,Chen:2018xof}, helicity~\cite{Alexandrou:2018pbm,Lin:2018pvv}, and transversity quark PDFs~\cite{Alexandrou:2018eet,Liu:2018hxv} have been calculated through the systematic procedure described in previous sections. Compared to earlier calculations without lattice renormalization~\cite{Lin:2014zya,Alexandrou:2015rja,Chen:2016utp,Alexandrou:2016jqi}, the updated results show significantly better agreement with PDFs from global analysis of high-energy scattering data.
	
	To illustrate the results, we take a recent calculation of the proton isovector quark helicity distribution by the LP$^3$ collaboration~\cite{Lin:2018pvv} as an example. The calculation was carried out with valence clover fermions on an ensemble of 884 configurations with lattice spacing $a=0.09$ fm, box size $L\approx 5.8$ fm, pion mass $M_\pi\approx 135$ MeV, and $N_f=2+1+1$ (degenerate up/down, strange and charm) dynamical flavors of highly improved staggered quarks (HISQ)~\cite{Follana:2006rc} generated by the MILC collaboration~\cite{Bazavov:2012xda}. Gaussian momentum smearing~\cite{Bali:2016lva} was used for the quark field to reach large proton boost momenta with $\vec{P}=\{0,0,n{2\pi\over L}\}$ and $n\in\{10,12,14\}$, corresponding to 2.2, 2.6, and 3.0 GeV, respectively. 1-step hypercubic smearing for the gauge link was used to suppress the discretization effects.
	
\begin{figure}[b]
		\centering
	\includegraphics[width=.6\textwidth]{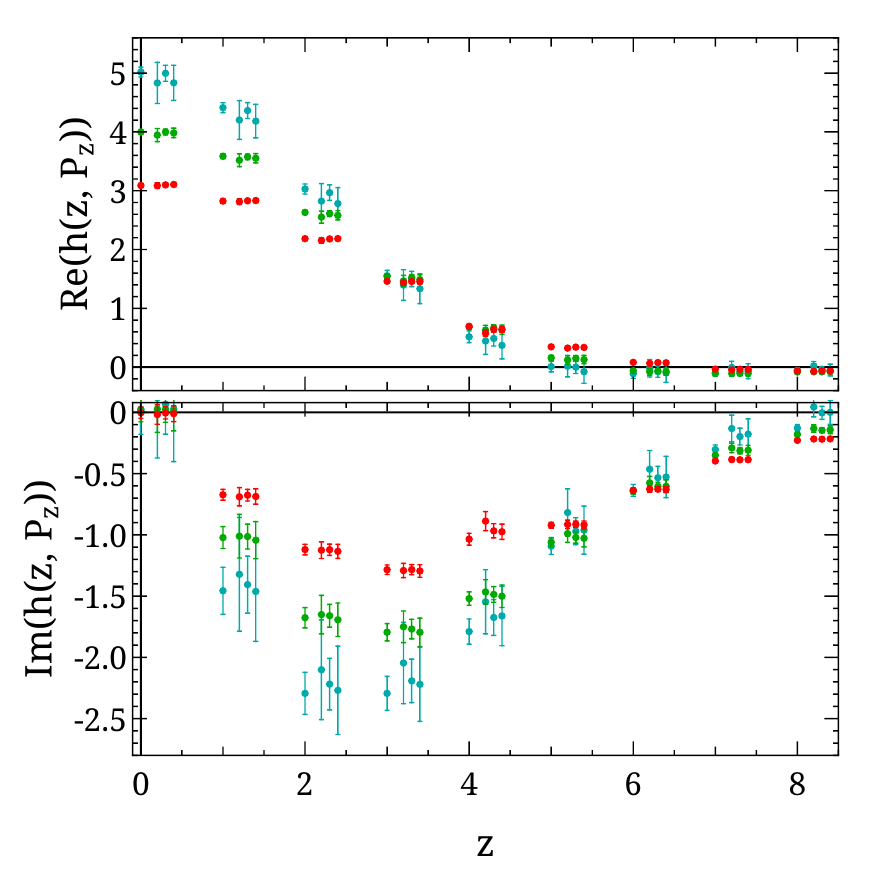}
	\caption{
	The real (top) and imaginary (bottom) parts of the bare proton matrix elements of $\tilde{O}_{\gamma^z\gamma_5}(z)$ for the iso-vector case at momentum $P^z=$2.2, 2.6 and 3.0~GeV, as indicated by red, green and blue points respectively. At a given positive $z$ value, the data are slightly offset horizontally to show different ground-state extraction strategies. From left to right: two-simRR using all $t_\text{sep}$, two-simRR using the largest five $t_\text{sep}$'s, two-sim using the largest four $t_\text{sep}$'s, and two-sim using the largest three $t_\text{sep}$'s. All four different fits yield consistent results.}
		\label{fig:bareME-tsep}
	\end{figure}

	The bare proton matrix elements of $\tilde{O}_{\gamma^z\gamma_5}(z)$ were calculated with six source-sink separations, $t_{\rm sep}\in\{0.54, 0.72, 0.81, 0.90, 0.99, 1.08\}$ fm with the number of measurements \{16, 32, 32, 64, 64, 128\}k, respectively. To control the excited-state-contaminations, four different two-state fitting strategies~\cite{Bhattacharya:2013ehc} were employed and consistent proton matrix elements were obtained, as shown in \fig{bareME-tsep}.
			
\begin{figure}[htbp]
		\centering
		\includegraphics[width=.6\textwidth]{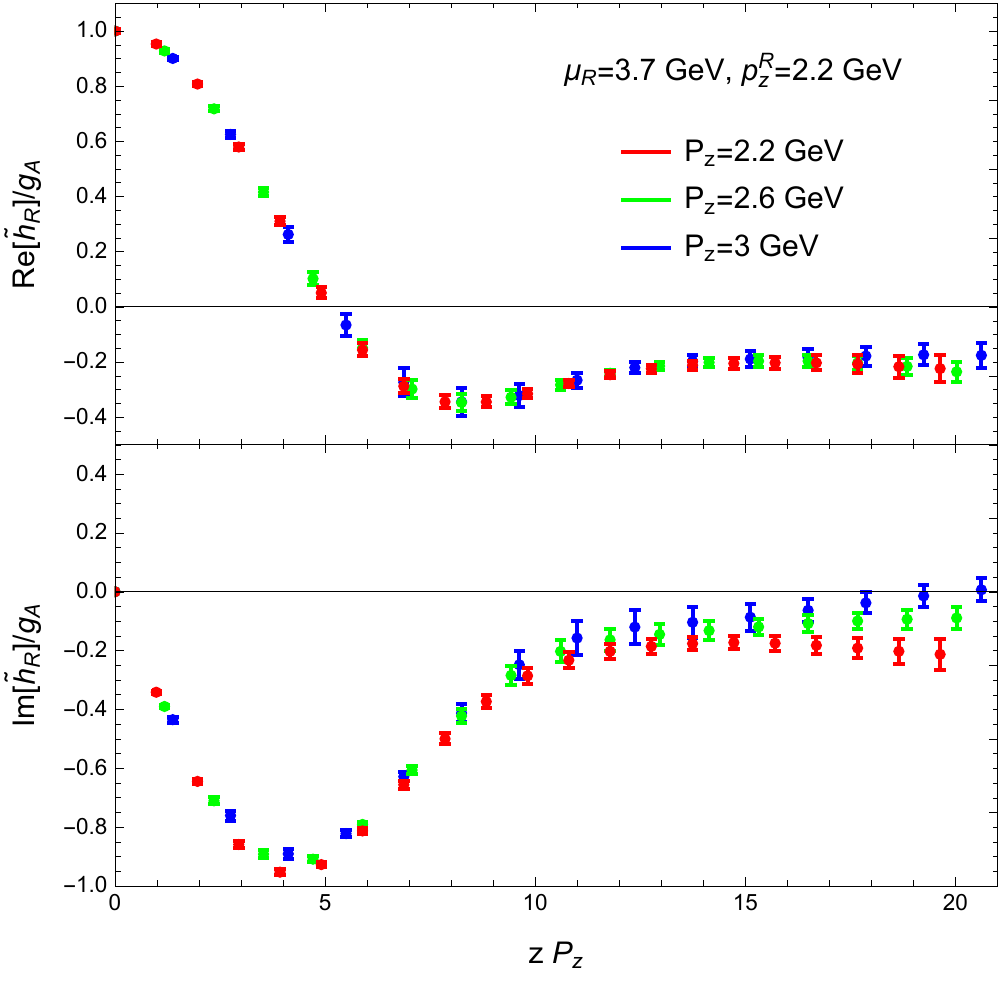}
		\caption{The real (top) and imaginary (bottom) parts of the renormalized proton matrix elements of $\tilde{O}_{\gamma^z\gamma_5}(z)$ as functions of $zP_z$, at renormalization scales
			$\mu_R=3.7$ GeV, and $p_z^R =2.2$ GeV.
		} \label{fig:ZME}
	\end{figure}

	The RI/MOM renormalization factor $Z$ was calculated from the amputated Green's function of $\tilde{O}_{\gamma^z\gamma_5}(z)$ with minimal projection in the Landau gauge on the same lattice ensemble. The $z$-component of the external quark momentum $p_z^R$ ranges from 0 to 3.0 GeV, whereas two values (2.3 and 3.7 GeV) were chosen for the subtraction scale $\mu_R$. The $p_z^R$ and $\mu_R$ dependence in the final result of the PDF shall be cancelled out by the perturbative matching and continuum limit. For the calculation in Ref.~\refcite{Lin:2018pvv}, only one lattice spacing was used and the perturbative matching was at one-loop order, so there are still remnant $p_z^R$ and $\mu_R$ dependence after all the systematic corrections. Therefore, $p_z^R=2.2$ GeV, $\mu_R=3.7$ GeV were chosen as central values, while the uncertainty from the remnant $p_z^R$ and $\mu_R$ dependence was estimated by the variation of $p_z^R$ and $\mu_R$ with the available data and included in the final result. The renormalized proton matrix elements of $\tilde{O}_{\gamma^z\gamma_5}(z)$ are shown in ~\fig{ZME}.
	
	After renormalization, one can then perform a Fourier transform to obtain the RI/MOM quasi-PDF. Due to the finite number of lattice data points, the Fourier transform was done by interpolating the points and truncating at a maximal value of $|z|$ for the integration over $z$. To eliminate the unphysical oscillation originating from the truncation, the derivative method~\cite{Lin:2017ani} was used for the Fourier transform:
	\beq
		\Delta \tilde{q}^R(x,P^z,\mu_R,p_z^R)=\int_{-\infty}^\infty \ {dz\over 4\pi }\ 	\frac{ie^{ixP^zz}}{xP^z} \partial_z\tilde{h}_R(z,P^z,\mu_R,p_z^R)\,,
	\eeq
	which is equivalent to the direct Fourier transform under the premise that $\tilde{h}_R(z,P^z,\mu_R,p_z^R)\to0$ in the limit of $|z|\to\infty$. For the lattice data, truncation of the above integral at $|z_{\rm max}|$ is in effect a cutoff of long range correlations, which will introduce uncontrolled systematic uncertainty for the quasi-PDF at small-$x$. To estimate such uncertainty, the $|z_{\rm max}|$ was varied beyond 15$a$. Since $\partial_z\tilde{h}_R(z,P^z,\mu_R,p_z^R)$ is consistent with zero for $|z|>15a$, the uncertainty is very small compared to the other systematics. The final result was taken at $|z_{\rm max}|P^z\sim 20$.
	
\begin{figure}[htbp]
		\centering
		\includegraphics[width=.6\textwidth]{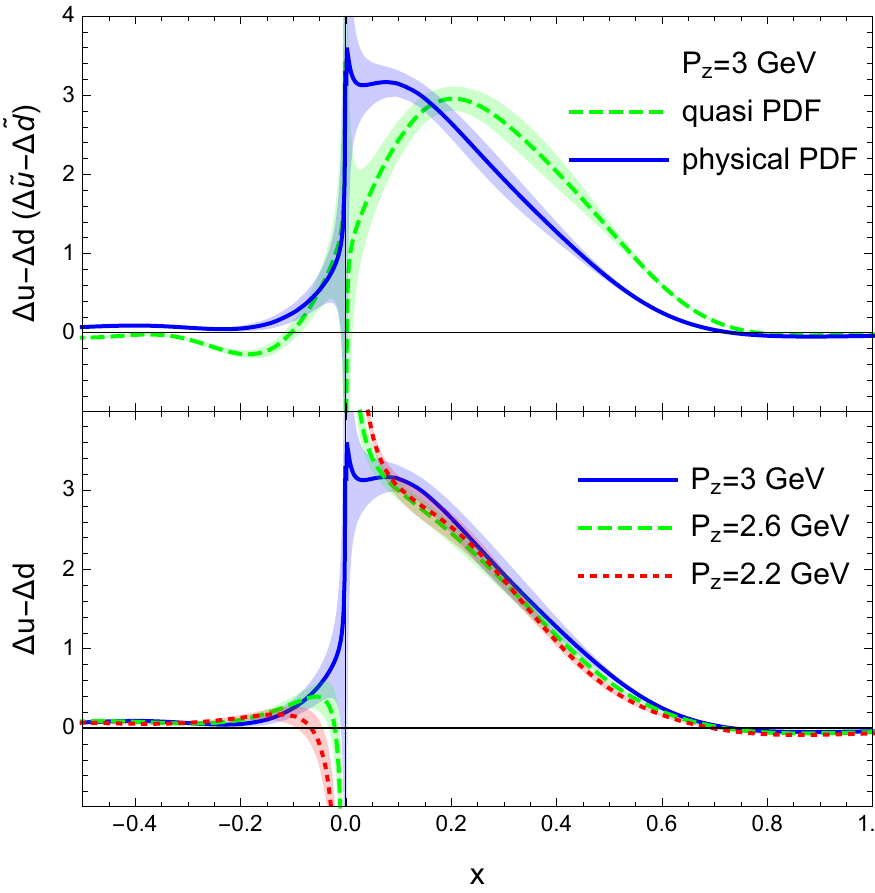}
		\caption{Top panel: quark helicity quasi-PDF in RI/MOM scheme at proton momentum $P^z=3.0$~GeV, $\mu_R=3.7$ GeV, $p_z^R=2.2$ GeV
			and resulting physical PDF in $\overline{\text{MS}}$ scheme at $\mu=3$~GeV. The error bands are statistical.
			The bottom panel shows the matched physical PDFs from all three proton momenta.} \label{fig:matchingPDF}
	\end{figure}

	The next step is to apply the mass and matching corrections, as shown in the upper panel of \fig{matchingPDF}. The mass correction is much smaller compared to the matching corrections. As one can see, the matching correction suppresses the quasi-PDF for $x>0.15$, while enhances it for $x<0.15$. Noticeably, a postive antiquark ($x<0$) helicity distribution was obtained, which is consistent with experiments. In the lower panel of \fig{matchingPDF}, the final results from all three different proton momenta are shown. For $x>0.1$, the
	differences among them are small, indicating that higher-twist effects are well suppressed in this region.
	In contrast, the central values for $x<0.1$ shift noticeably from $P^z=2.2$ to 3.0~GeV.
	
\begin{figure}[htbp]
		\centering
			\includegraphics[width=.6\textwidth]{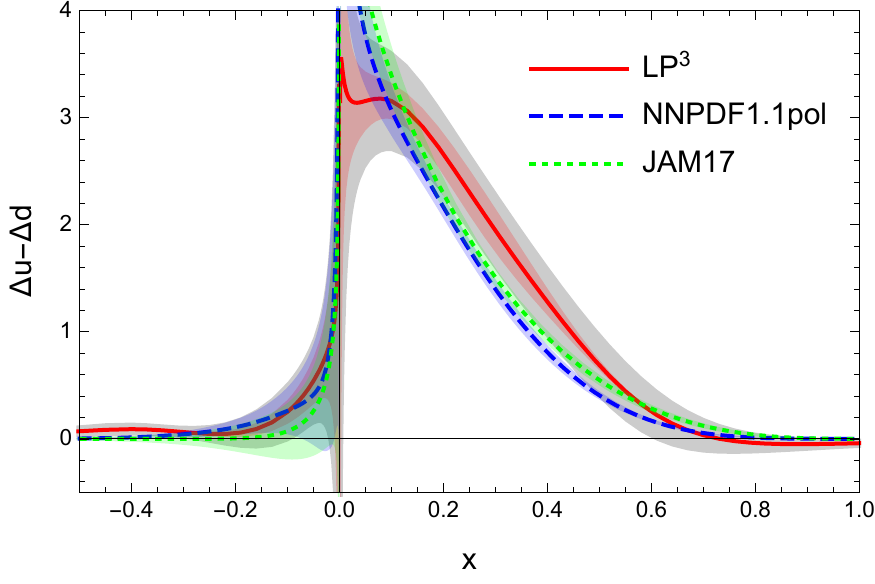}
			\caption{Comparison of the lattice result of proton isovector quark helicity PDF at $\mu=3.0$ GeV with fits by NNPDFpol1.1~\cite{Nocera:2014gqa} and JAM~\cite{Ethier:2017zbq}. The red band contains
				statistical error while the gray band also includes estimated systematics from finite lattice spacing, finite
				volume, higher-twist corrections, as well as renormalization scale uncertainties.
			} \label{fig:finalPDF}
	\end{figure}

		Finally, a comparison of the final result for $P^z=3.0$ GeV and two recent global analysis of the isovector quark helicity PDF by NNPDFpol1.1~\cite{Nocera:2014gqa} and JAM~\cite{Ethier:2017zbq} is shown in \fig{finalPDF}. The lattice result is consistent with both analyses within 1$\sigma$ for the moderate-$x$ region $0.1<x<0.7$. For $x$ close to 1, the prediction is limited by the finite lattice spacing, whereas the systematic uncertainty for $0<x<0.1$ is larger and not well under control. The antiquark distribution shows agreement with the experimental fits, but the uncertainties are also larger.
		
	In addition to the isovector quark helicity PDF, similar results have also been obtained for the unpolarized distribution~\cite{Chen:2018xof}. The remarkable agreement between the lattice calculation and global analyses of the PDF shows that the systematic corrections to the quasi-PDFs are improving the results in the right direction. This improvement has also been observed by the recent lattice calculations done by the ETM Collaboration~\cite{Alexandrou:2018pbm}.
	
\begin{figure}[htbp]
		\centering
		\includegraphics[width=.6\textwidth]{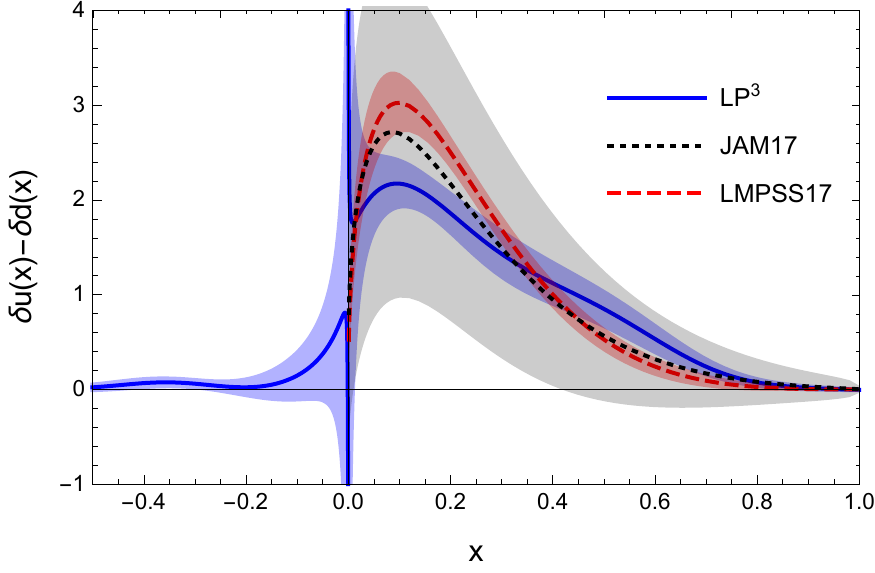}
		\caption{Comparison of the lattice result of proton isovector quark transversity PDF at $\mu=3.0$ GeV with fits by JAM17 and LMPSS~\cite{Lin:2017stx}.
		} \label{fig:transversePDF}
	\end{figure}

	Finally, it is worthwhile to mention that, the lattice calculations of the iso-vector transversity PDF with the LaMET approach, which has been done by the ETM~\cite{Alexandrou:2018eet} and LP$^3$~\cite{Liu:2018hxv} collaborations, have already reached a precision that is even better than the up-to-date experimental fits. \fig{transversePDF} shows a recent calculation using the same lattice ensemble by the LP$^3$ collaboration~\cite{Liu:2018hxv}, where the lattice result of the isovector transversity PDF was compared to the global analysis by JAM17~\cite{Lin:2017stx} and a combined fit (LMPSS~\cite{Lin:2017stx}) constrained by the lattice-average of the tensor charge $g_T$, which is the lowest moment of the transversity PDF. As one can see, the lattice result has much smaller uncertainties compared to global analysis. The combined analysis with constraints from lattice-averaged $g_T$ can significantly improved the precision, and is consistent with the LP$^3$ result within 2$\sigma$ in the positive $x$ region. More importantly, the lattice calculation can predict the anti-sea flavor asymmetry to be consistent with zero, which has been assumed by all global analyses so far. This calculation can provide an important constraint on the experimental fits of transversity PDFs.\\
	
	For a complete analysis of the systematics, one should also study: 
	1) Lattice discretization effects. As has been discussed in \sec{ren}, the renormalized quasi-PDF should have a continuum limit as $a\to0$. At finite $a$, the discretization effects could be of the order $O(P_z^2a^2, \mu_R^2a^2, a^2(p_z^R)^2)$. It is highly desirable to eliminate such effects by calculating at several different spacings and extrapolating to the continuum. 2) Matching at higher loop orders. As shown in \fig{matchingPDF}, the next-to-leading order matching brings considerable corrections to the renormalized quasi-PDF, so it is necessary to consider the next-to-next-to-leading order matching to examine the convergence of perturbative corrections. 3) Finite volume effects. Since the lattice calculations are done in a finite volume, the periodic boundary condition can lead to significant finite volume effects when the extension of the nonlocal operator is comparable to half of the size of the lattice, as was found from a one-loop study of the scalar current-current correlator~\cite{Briceno:2018lfj}. This will force a truncation of the long range correlation for the quasi-PDF, thus leading to uncontrolled systematics in the small-$x$ distribution. Meanwhile, to extract the physical matrix elements, one should also do the calculation with different volumes and extrapolate to the infinite volume limit. Nevertheless, with large hadron momentum, the finite volume effects can be contained as the short range correlations dominate.
	4) Resummation of matching corrections in certain kinematic regions. The corrections in the end-point regions $x\to0$ and $x\to1$ can be more significant than the tree-level matching (i.e. no matching correction) result, then resummation will become necessary.
	
	To improve the prediction for small-$x$ distributions, one will ultimately rely on having finer lattice spacings so that within the window $a^{-1}\gg P^z \gg \Lambda_{\rm QCD}^{-1}$, one can obtain large enough $P^z$ to suppress $O(\Lambda_{\rm QCD}^2/(xP^z)^2)$ effects for a wider range of $x$ while keeping the discretization effects small. Nevertheless, it will be helpful to calculate the higher-twist corrections with proper renormalization and matching on lattice, though it requires more sophisticated techniques. 

	\section{Other developments of LaMET}
	\label{sec:other}
	
	Currently, the lattice calculation of PDFs is limited to the isovector cases. To separate the quark flavors, one has to include the mixing with gluons, which will require us to calculate the gluon quasi-PDF~\cite{Ji:2013dva} on lattice. A first calculation of the matrix elements of the equal-time correlators for the gluon quasi-PDF on lattice has already been carried out~\cite{Fan:2018dxu}. Just like the quark case, the gluon quasi-PDF also needs to undergo renormalization and perturbative matching. It has been proven that the gluon quasi-PDF can also be multiplicatively renormalized in coordinate space~\cite{Zhang:2018diq,Li:2018tpe}, and its matching along with the singlet quark quasi-PDF has been studied at one-loop order in the transverse momentum cutoff and $\overline{\rm MS}$ schemes~\cite{Wang:2017qyg,Wang:2017eel}. A nonperturbative renormalization of gluon quasi-PDF on lattice can also be carried out in the RI/MOM scheme~\cite{Zhang:2018diq}, and with the matching kernels calculated in perturbative QCD~\cite{Wang:2019xxx}, one can expect to see a first lattice calculation of gluon PDF as well flavor separated or singlet quark PDFs with the LaMET approach.\\
	
	In addition to the one-dimensional parton distributions, we can also extend our study to the transverse structures of the hadron.
	
	The GPD is defined as the off-forward hadron matrix elements of the same light-cone correlator as the PDF. Its dependence on the momentum transfer of the hadron will provide information of the distribution of partons in the transverse coordinate space, similar to form factors. To calculate GPDs from lattice with LaMET, one can start from the quasi-GPD that is defined with the same equal-time correlator $\tilde{O}_\Gamma(z)$ for the quasi-PDF. Since the renormalization and matching for the quasi observables are determined by the UV or short distance properties of the operator itself, it is natural to apply the results from \secs{ren}{match} for the quasi-GPDs. In practice, the quasi-GPD can be renormalized with the same factors for the quasi-PDF, but the matching needs to be re-derived. The matching kernels for the quasi-GPDs in a transverse momentum cutoff scheme has been derived in literature~\cite{Ji:2015qla,Xiong:2015nua}, while the matching kernels for the RI/MOM quasi-GPD~\cite{Liu:2018xxx} will be the necessary ingredient for the lattice calculations. As a special limit of GPD, the one-loop matching kernel of the quasi distribution amplitudes in the RI/MOM scheme is already available~\cite{Liu:2018tox}.
	
	On the other hand, the TMDPDFs describe the structure of hadrons in the transverse momentum ($q_T$) space. They are also defined from light-cone correlators, except that it includes a staple-shaped Wilson line that extends to light-cone infinity which leads to the so called ``rapidity'' divergences, and a soft factor subtraction is needed to cancel these divergences and give phenomenologically meaningful TMDPDFs in factorization theorems~\cite{Collins:2011zzd}. To calculate the TMDPDFs from lattice with LaMET, one should also construct quasi distributions for both the staple-shaped Wilson line operator as well as the soft factor. While the former can be naively generalized from the quasi-PDF case and perturbatively matched onto the light-cone correlator, the latter faces intrinsic obstacle as it involves Wilson lines in two distinct light-cone directions that cannot be related to an equal-time correlator through a single Lorentz boost.
	
	The construction of quasi observables for the TMDPDF and their perturbative matching has been studied in Refs.~\refcite{Ji:2014hxa,Ji:2018hvs,Ebert:2018xxx}. It is found that for certain definitions of the quasi soft factor, a perturbative matching onto to the TMDPDF exists at one-loop order~\cite{Ji:2014hxa,Ebert:2018xxx} for $q_T\sim\Lambda_{\rm QCD}$. Recently, a method was developed to determine the nonperturbative anomalous dimension for the TMDPDF evolution in the rapidity scale from lattice QCD~\cite{Ebert:2018gzl}. Besides being a novel target for lattice QCD, this will also enable important improvements to experimental fits of the TMDPDFs.
	After all, it is necessary to continue pursuing the lattice calculation of TMDPDFs with the goal of ultimately overcoming other complications that arise in these challenging observables.

	\section{Conclusion}
	\label{sec:concl}
	
	In this paper, we have described the formalism of LaMET and shown how it can be used as a general approach to calculate parton physics in lattice QCD. We have also provided details of the systematic procedure of calculating light-cone PDFs with this approach, which includes nonperturbative lattice renormalization, perturbative matching, as well as mass and higher-twist corrections. This systematic procedure has been applied to the recent lattice calculations of proton isovector quark PDFs for the unpolarized~\cite{Alexandrou:2018pbm,Chen:2018xof}, helicity~\cite{Alexandrou:2018pbm,Lin:2018pvv}, and transversity cases~\cite{Alexandrou:2018eet,Liu:2018hxv} at physical pion mass, and remarkable agreements with global fits from experiments have been reached. Especially, the precision of the transversity PDF is better than the up-to-date global analysis, and lattice calculation is capable of predicting the anti-sea distributions that have been assumed to be zero in experimental fits, which will have significant impact on phenomenology.
	
	Such progress shows a promising sign that the systematic corrections in LaMET are working in the right direction to improve the prediction of light-cone PDFs, and it is worthwhile to investigate the other systematics in this program in order to reach a better precision with current computing resources. In the short term, lattice calculations should aim at providing complementary information of the PDFs that are not well constrained from current experiments. In the long term, with increased computing power and improved systematics, one can evision precision calculations. Since the multi-dimensional hadron structure will be among the top scientific goals at the electron-ion collider~\cite{Accardi:2012qut}, the frontier of QCD in the next decade, it will be exciting to see the interplay between lattice QCD and high-energy experiments providing accurate description of hadronic structures and input for Standard Model calculations.

	\section*{Acknowledgments}
	
	We thank A. Alexandru, J.-W. Chen, W. Detmold, T. Drapper, M. Ebert, Y. Hatta, T. Ishikawa, T. Izubuchi, X. Ji, L. Jin, I. Kanamori, R. Li, C.-J. D. Lin, H.-W. Lin, K.-F. Liu, Y.-S. Liu, S. Mondal, A. Sch\"{a}fer, I. Stewart, P. Sun, R.~S. Sufian, W. Wang, X. Xiong, J. Xu, Y. Xu, Y.-B. Yang, F. Yuan, J.-H. Zhang, Qi-An Zhang, Rui Zhang and Shuai Zhao for fruitful discussions and collaborations. We are also indebted to the englightening discussions with G. Bali, V. M. Braun, J. Chang, Y.-T. Chien, K. Cichy, M. Constantinou, M. Engelhardt, J. Green, Y. Jia, K. Jansen, J. Karpie, H.-N. Li, S. Liuti, Y.-Q. Ma, A. Manohar, A. Metz, B. Mistlberger, C. Monahan, D. Murphy, J. Negele, K. Orginos, P. Petreczky, A. Pochinsky, A. Prokudin, J. Qiu, F. Ringer, P. Shanahan, F. Steffens, S. Syritsyn, M. Wagman, X.-N. Wang, Y. Yin, H.-X. Zhu.
	
	This work is supported by the U.S. Department of Energy, Office of Science, Office of Nuclear
	Physics, from DE-SC0011090 and within the framework of the TMD Topical Collaboration. YZ is also supported in part by the MIT MISTI program and by the Munich Institute for Astro- and Particle Physics (MIAPP) of the DFG cluster of excellence ``Origin and Structure of the Universe''. YZ is grateful for the hospitality of Shanghai Jiao-Tong University during this stay at SKLPPC, T-D Lee Institute, and National Chiao-Tung University where the collaborations were made possible.

\bibliographystyle{ws-ijmpa}
\bibliography{lametrefs}
\end{document}